\documentclass[sigconf]{acmart}

\usepackage{svg}
\usepackage{listings}
\usepackage{graphicx}
\usepackage{subcaption}
\usepackage[normalem]{ulem}
\useunder{\uline}{\ul}{}

\begin{document}
  
\title{Comparative analysis of real bugs in open-source Machine Learning projects - A Registered Report}


%\author{Dung Lai}
% \affiliation{%
%   \institution{Deakin University}
%   \department{Applied Artificial Intelligence Inst.}
%   \streetaddress{Locked Bag 20000}
%   \city{Geelong}
%   \state{VIC}
%   \country{Australia}
%   \postcode{3220}
% }
%\email{tuan.lai@deakin.edu.au}
%\author{Andrew J. Simmons}
%\orcid{0000-0001-8402-2853}
% \affiliation{%
%   \institution{Deakin University}
%   \department{Applied Artificial Intelligence Inst.}
%   \streetaddress{Locked Bag 20000}
%   \city{Geelong}
%   \state{VIC}
%   \country{Australia}
%   \postcode{3220}
% }
%\email{a.simmons@deakin.edu.au}
\author{Tuan Dung Lai, Anj Simmons, Scott Barnett, Jean-Guy Schneider, Rajesh Vasa}
% \affiliation{%
%   \institution{Deakin University}
%   \department{Applied Artificial Intelligence Inst.}
%   \streetaddress{Locked Bag 20000}
%   \city{Geelong}
%   \state{VIC}
%   \country{Australia}
%   \postcode{3220}
% }
\email{{tuan.lai, a.simmons, scott.barnett,jeanguy.schneider,rajesh.vasa}@deakin.edu.au}
%\author{Iman Avazpour}
%\email{iman.avazpour@deakin.edu.au}

%\affiliation{%
%  \institution{Deakin University}
%  \department{Applied Artificial Intelligence Inst.}
%  \streetaddress{Locked Bag 20000}
%  \city{Geelong}
%  \state{VIC}
%  \country{Australia}
%  \postcode{3220}
%}
%\affiliation{%
%  \institution{Deakin University}
%  \department{School of Information Technology}
%  \streetaddress{Locked Bag 20000}
%  \city{Geelong}
%  \state{VIC}
%  \country{Australia}
%  \postcode{3220}
%}
%
%\author{Jessica Rivera-Villicana}
% \affiliation{%
%   \institution{Deakin University}
%   \department{Applied Artificial Intelligence Inst.}
%   \streetaddress{Locked Bag 20000}
%   \city{Geelong}
%   \state{VIC}
%   \country{Australia}
%   \postcode{3220}
% }
%\email{jessica.riveravillicana@deakin.edu.au}
%
%\author{Shangeetha Sivasothy}
% \affiliation{%
%   \institution{Deakin University}
%   \department{Applied Artificial Intelligence Inst.}
%   \streetaddress{Locked Bag 20000}
%   \city{Geelong}
%   \state{VIC}
%   \country{Australia}
%   \postcode{3220}
% }
%\email{s.sivasothy@deakin.edu.au}
%
%\author{Rajesh Vasa}
%\email{rajesh.vasa@deakin.edu.au}
\affiliation{%
  %\institution{Deakin University}
  \department{Applied Artificial Intelligence Inst., Deakin University}
  %\streetaddress{Locked Bag 20000}
  \city{Geelong}
  \state{VIC 3220}
  \country{Australia}
  \postcode{3220}
}
% \email{{tuan.lai, a.simmons, scott.barnett, jessica.riveravillicana, iman.avazpour rajesh.vasa, rajesh.vasa}@deakin.edu.au}

\renewcommand{\shortauthors}{Lai et al.}

\begin{abstract}

\textbf{Background:} 
Machine Learning (ML) systems rely on data to make predictions, the systems have many added components compared to traditional software systems such as the data processing pipeline, serving pipeline, and model training.
Existing research on software maintenance has studied the issue-reporting needs and resolution process for different types of issues, such as performance and security issues.
%Existing research on software maintenance has pointed out the need for alerting and detecting issues which occur unexpectedly during production due to the change in the environment, change in stream of data, and the change in the software components.
However, ML systems have specific classes of faults, and reporting ML issues requires domain-specific information. Because of the different characteristics between ML and traditional Software Engineering systems, we do not know to what extent the reporting needs are different, and to what extent these differences impact the issue resolution process.
%whether Software Engineering reporting techniques and literature can be applied for ML.

% Robustness in software engineering cannot be guaranteed for machine learning systems because it needs monitoring and testing which requires a decent reporting to maintain and fix. Reporting template in open-source applied AI projects do not contain necessary information required to fix ML specific bugs. Because of the different characteristics between machine learning and software engineering systems, we do not know to what extent its reporting techniques and literature can be applied for machine learning.

% Recent studies in defects in software context focus on defect classification, automatic defect assignment and defect severity prediction, researches in defect report primarily focus on understanding defect report contents, predicting usefulness of reported information, predicting defect characteristics and building defect reporting tools. 

\textbf{Objective:} 
Our objective is to investigate whether there is a discrepancy in the distribution of resolution time between ML and non-ML issues and whether certain categories of ML issues require a longer time to resolve based on real issue reports in open-source applied ML projects. We further investigate the size of fix of ML issues and non-ML issues.

% We also want to measure the size of fix required for ML and non-ML issues and also between ML categories.
\textbf{Method:} 
We extract issues reports, pull requests and code files in recent active applied ML projects from Github, and use an automatic approach to filter ML and non-ML issues. We manually label the issues using a known taxonomy of deep learning bugs. We measure the resolution time and size of fix of ML and non-ML issues on a controlled sample and compare the distributions for each category of issue.

%in a controlled environment then finally use statistical analysis to compare the differences and describe any correlation.

\end{abstract}





\maketitle

\graphicspath{{figures/}}
\section{Introduction}
\label{sec:introduction}

\begin{figure}[htbp]
  \centering
  \includegraphics[width=\columnwidth]{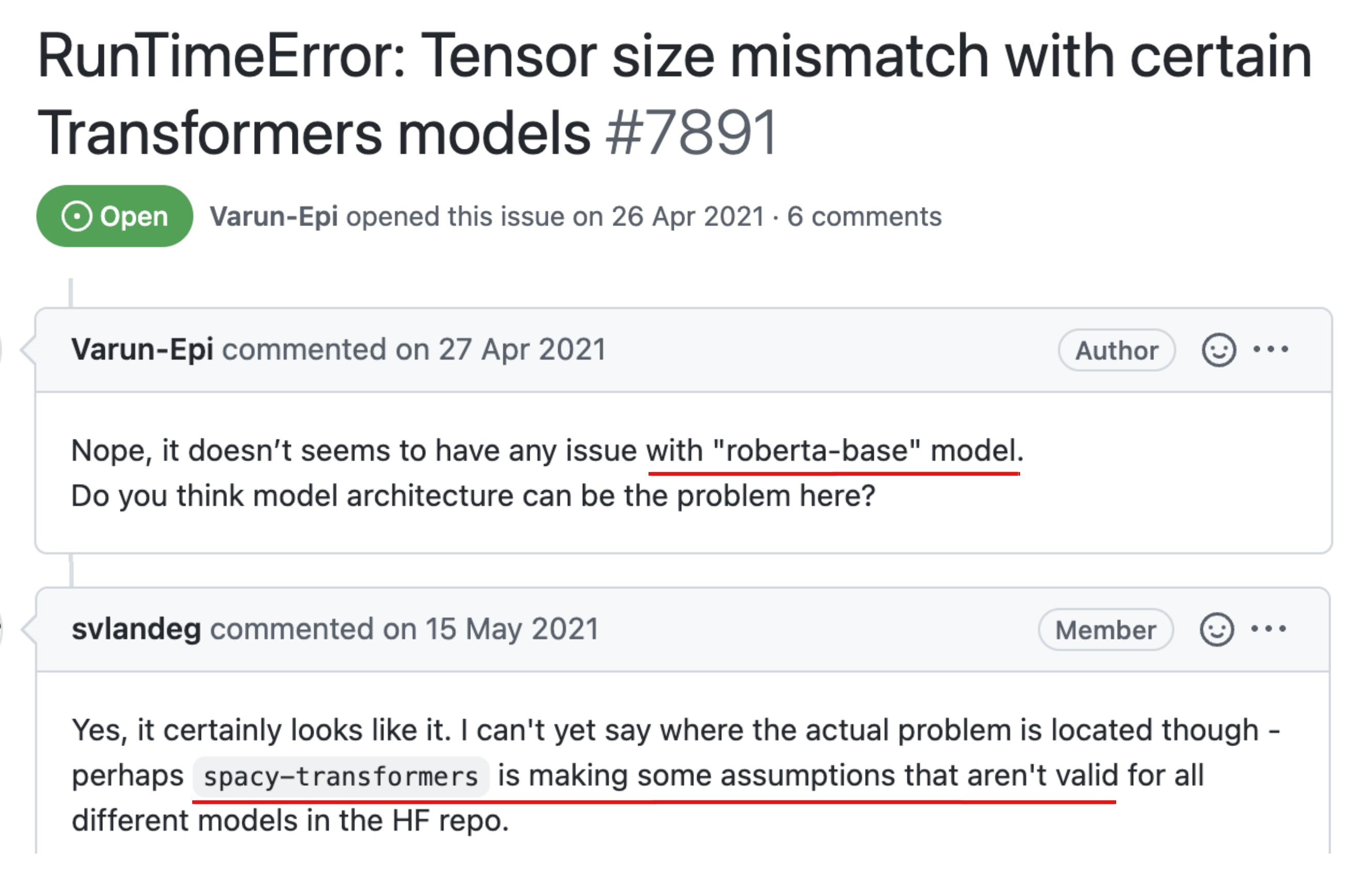}
  \caption[An open ML issue on spaCy related to the model architecture found during model training.]{An open ML issue on spaCy\footnotemark~related to the model architecture found during model training.}
  \label{fig:issue}
\end{figure}

Software issues are well studied and best practices for reporting issues is well known~\cite{ghanavati2020memory, davies2014bug, bhattacharya2013empirical, tan2014bug, chou2001empirical,lal2012comparison, li2006have}. In this report, we use \textit{ML issues} to refer to tickets raised in open-source ML projects that indicate an ML-specific problem in the ML code of the software (as opposed to non-ML issues related to the system surrounding the ML code, such as configuration, data collection, and data verification components). The ML issue shown in \autoref{fig:issue} shows an issue that occurs when training a model based on specific transformer models, i.e., the issue arises only when using certain ML models. This category of issue is specific to ML. The implication of these ML issues on resolution time and size of the fixes is unknown. By investigating ML issues new techniques can be devised for efficient and effective ML issue reporting and resolution strategies.  

% However, research also shows that best practice depends on the domain with specific techniques developed for issues in security(CITE), performance(CITE) and user interfaces(CITE). This study investigates if the same is true of machine learning related issues. 

% Thus, reliability of ML is achieved through efficient and effective issue detection and resolution strategies (CITE).

Existing methods to maintain and monitor traditional software systems are not sufficient for systems that have machine learning components because of the non-stationary nature of the system which involves data and models components that evolve and changed constantly \cite{parker2015detecting}. Mounting evidence also shows that ML has unique failure modes~\cite{humbatova2020taxonomy}, no known techniques for robustness guarantees (ML will eventually fail), and gradual degradation occurs due to data shift~\cite{Wang2018ADrift}. ML specific testing strategies help but are insufficient for guaranteeing bug free software -- failure cases will be missed during testing.  However, fixing issues in ML involve an exploration process to investigate i) the training process, ii) modelling assumptions, and iii) changes in the data. Studying the impact of issues in ML has implications on a) issue assignment, b) issue reporting templates, and c) task prioritisation. 

% issue assignment: \cite{anvik2011reducing}

Current research on ML issues focuses on bug classification \cite{sun2017empirical, humbatova2020taxonomy, thung2012empirical}, root cause analysis \cite{shen2021comprehensive, zhang2019empirical, zhang2020empirical, islam2019comprehensive, zhang2018empirical, chen2022toward} and defect detection \cite{wardat2022deepdiagnosis, liu2021detecting, yan2021exposing}. This research provides an understanding of the characteristics of ML issues but does not compare to our existing body of knowledge on software issues. Without drawing a direct comparison to prior knowledge we cannot assess if existing best practices apply for ML issues. Specifically we do not know if ML issues follow the same distribution for i) resolution time or ii) fix size. %From our preliminary investigation into ML issues we hypothesise that there are differences between ML and non-ML issues but whether this has an impact on resolution time or fix size is unknown. Hence, the purpose of this study.  

% Prior work in ML issues has not investigated if ML issues are similar to non-ML issues.

% Whether these characteristics impact the reporting or resolution time of ML issues is unknown.  

% There is also multiple terms to refer to issues in the literature~\cite{zhang2020empirical,sun2017empirical,humbatova2020taxonomy,arya2019analysis, islam2019comprehensive}.  Prior work has not investigated whether our understanding of software issues (in terms of resolution time and information emergence)  

% Our research intends to addresses the following gaps: i) no insight into the effectiveness of ML defect reports, and ii) no empirical evidence for frequency of ML defects categories.

We investigate the following: (1) frequency of ML issue categories to empirically evaluate an existing taxonomy~\cite{humbatova2020taxonomy}; (2) the differences between ML and non-ML issues in terms of resolution time and size of fix; and (3) compare the different ML issue categories with regards to resolution time and size of fix. Our research is intended to guide future work into ML issues.  

% The expected research outcomes from this work include 1) an empirically derived dataset of annotated ML issues, 2) an evaluation of the frequency of ML issue categories in open-source projects, 3) an empirical comparison between ML and non-ML issues, and 4) an empirical comparison between ML issue categories. This research has the potential to open up new research avenues specific to ML issues.

\section{ML Issue Lifecycle}

\begin{figure}
    \centering
    \includegraphics[width=0.45\textwidth]{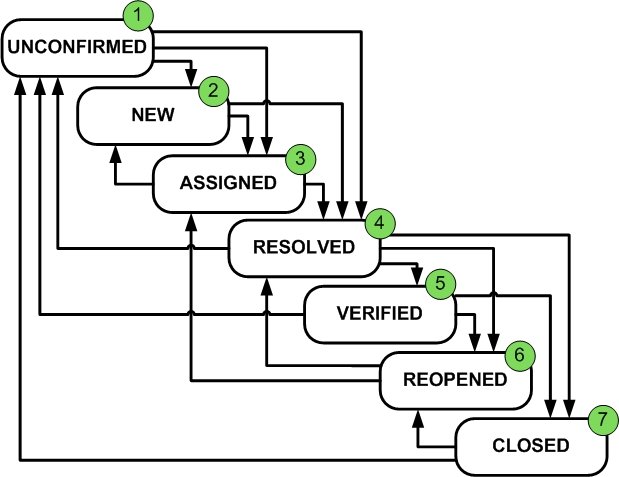}
    \caption{Bugzilla issue life cycle \cite{gegick2010identifying}}
    \label{fig:buglifecycle}
\end{figure}
%\footnotetext{http://www.atlassian.com/software/jira/ }

The life cycle of a software issue is well understood~\cite{gegick2010identifying}. An example life cycle for the Bugzilla issue tracker is shown in \autoref{fig:buglifecycle}. Our overall hypothesis is that ML issues will take logner to fix than non-ML issues. Below we motivate how delays for ML issues can occur at each stage of the issue lifecycle. Issues that take an unexpected amount of time to close impact project schedules and increases maintenance costs.%\cite{}.  

\footnotetext{https://github.com/explosion/spaCy}

\textit{Unconfirmed, New.} Issues raised for a software system occur when a user experiences a system failure. Users fill out an issue report that documents the failure in a ticketing system to be assigned to a developer. Project teams need a thorough understanding of the application and the category of issue to be able to estimate time to create a fix. However, ML issues have specific defect categories~\cite{humbatova2020taxonomy}. In some cases, the only symptom of these defects is that the prediction is incorrect \cite{Florian2021}. This requires exploration time to understand the root cause. Whether the categories of ML issues impacts resolution time is unknown.

\textit{Assigned.} Knowing who to assign an ML issue to is challenging as ML issues occur in a) the architecture of the model, b) training pipelines, c) choice of hyperparameters, d) serving pipelines, and d) assumptions about inputs. Different skills are needed to address an ML issue depending on the location including MLDevOps, Data Engineers, Software Engineers, and ML experts. Assessing the right developer to resolve an ML issue influences the time to create a resolve. 

\textit{Resolved.} Once a developer has been assigned an issue they attempt to locate, and resolve the issue. To resolve an ML issue like the one in \autoref{fig:issue} understanding is required of the different model architectures and the training process that causes the error. Resolving ML issues requires time to prepare, traing, and validate an ML model. 

\textit{Verified.} Verifying of ML is an on-going research area~\cite{xiang2018verification}. The difficulties of verifying if an ML issue has been resolved lie in understanding the unbounded function learned by the machine -- the limitations of the machine cannot be inspected by a human. We hypothesise that the challenges with verifying an ML issue would contribute to a longer resolution time than non-ML issues.  

\textit{Closed, Re-Opened.} ML issues resolved by retraining involve no code or configuration changes (i.e., only model versions and training data are updated). This requires additional infrastructure outside of the source code repository indicating how the training strategy (i.e., hyperparameters, and training data) addressed the issue for replication.

\section{Related Work}

Research on issues in ML applications has focused on categorising bugs and building taxonomies \cite{sun2017empirical, humbatova2020taxonomy, thung2012empirical}, finding root causes \cite{zhang2019empirical, zhang2020empirical, islam2019comprehensive, zhang2018empirical}, ranking bugs by frequency and severity \cite{zhang2020empirical, thung2012empirical, zhang2019empirical, islam2019comprehensive}, extracting common fix patterns \cite{sun2017empirical, zhang2020empirical}, and measuring resolution time \cite{sun2017empirical, thung2012empirical}. Some class of ML defects are more severe than others and requires a extended time to fix \cite{sun2017empirical}. In 2017, \citet{sun2017empirical} conducted an empirical study on real bugs in ML programs from 329 ML issues from 3 large open-source ML projects from Github and found that nearly 70\% of bugs are fixed within one month, certain types of defects are more severe than others and require a extended time to fix \cite{sun2017empirical}. However, they did not compare the empirical statistics against a baseline for non-ML issues. In 2019, \citeauthor{arya2019analysis} used supervised learning to extract common topics from 15 complex issues reports and its comments from 3 open-source ML projects \citet{arya2019analysis}. The aim of their study is not to understand ML specific issues but to generalise these topics to software projects. In 2020, \citeauthor{humbatova2020taxonomy} analysed GitHub issues/PRs/commits and Stack Overflow discussion to propose a taxonomy of deep learning faults and conducted interviews with practitioners to rank categories of bugs based on severity and effort required \cite{humbatova2020taxonomy}. However, the severity and effort estimates for each category are based on interviews rather than an empirical analysis of issues.

\section{Research Questions}
To guide the research and investigation we propose the following research questions. 

% \textbf{RQ1. What topics are mentioned when documenting an ML issue?} 
\textbf{RQ1. What is the frequency of ML issue categories in open-source projects?} 
Our goal is to partially replicate the taxonomy created by  \citet{humbatova2020taxonomy}. Replication will add to the empirical evidence in support of the taxonomy and inform research direction, e.g., what are the common ML issues to focus on for auto-repair techniques? We will attempt to apply the taxonomy in \cite{humbatova2020taxonomy} to classify our ML issues and identify which categories are most frequent and whether there are any categories missing (which we will classify as `other' if present). The classification of ML issues is also required for our later research questions.  

\begin{itemize}
    \item[H1.1] The categories of ML issues in our dataset will occur with similar frequency to that observed by \citet{humbatova2020taxonomy}. However, there may be minor differences as a result of different sampling strategies and analysing a broader class of ML projects than just deep learning.
    \item[H1.2] No new categories will emerge from our dataset of ML issues as the original taxonomy \cite{humbatova2020taxonomy} was created from investigating Stack Overflow, Github and interviews.
\end{itemize}

% Addressing this research question has implications for automatic classification of defects\cite{pandey2017automated}.

% todo
% Previous research has identified taxonomy of ML bugs based on interview answers from industry practitioners \cite{humbatova2020taxonomy}, In our research, we plan to investigate what information are reported based on empirical evidence.

% Qualitatively analyse ML issues report and draw comparison with non-ML issues.

% \textbf{RQ2. How do topics that emerge after the initial issue report differ between ML and non-ML issues?} 

% The purpose of this research question is to discover what information is required after the initial issue report is created. Our intention is to see if ML characteristics of training strategy, modelling assumption, or choice of algorithm impact the information required to resolve an issue.  

% \begin{itemize}
%     \item[H2.1] ML issues will have more information required after the initial issue report than non-ML issues. 
%     \item[H2.2] ML issues will have different information emerge after the initial issue report than non-ML issues. 
%     \item[H2.3] Missing ML specific information (such as training strategy, modelling assumptions, and data distribution) required for fixing an ML issue will emerge after the initial issue report. The information required is not available when the failure occurs. 
% \end{itemize}

% Using topic modelling on issue title, descriptions and comments.

\textbf{RQ2. How does the distribution of ML and non-ML issues compare in terms of resolution time and size of fix?}
The purpose of this research question is to investigate if the characteristics of ML impact the resolution time and size of a fix. Our goal in addressing this research question is to investigate if:   

% We plan to compare 2 dimensions:
% resolution time
% size of fix
% With RQ2 we  find that ML issues take longer to resolve than non-ML issues. We expect to find a left skewed distribution with a statistical difference with the distribution of non-ML issues. 

\begin{itemize}
    \item[H2.1] There will be no difference in the size of the fix between ML and non-ML issues. 
    \item[H2.2] ML issues take longer to resolve than non-ML issues. We hypothesise that attributes of ML will require longer resolution times due to i) training strategies, ii) modelling assumptions and iii) location of defect (i.e., in model architecture, training strategy, hyper-parameter configuration, or code implementation). 
\end{itemize}

\textbf{RQ3. How does the distribution of different ML issue categories compare in terms of resolution time and size of fix?}

% With RQ2 we hope to find that ML issues take longer to resolve than non-ML issues. We expect to find a left skewed distribution with a statistical difference with the distribution of non-ML issues. 

\begin{itemize}
    \item[H3.1] Training data and training process are the two categories of ML issues that take the \textbf{longest time} to resolve because \citeauthor{humbatova2020taxonomy} found that they have the highest severity and effort required to fix according to their survey and interview with practitioners \cite{humbatova2020taxonomy}.
    \item[H3.2] Training data and training process are the two categories of ML issues that take the biggest \textbf{size of fix} to resolve because \citeauthor{humbatova2020taxonomy} found that they have the highest severity and effort required to fix according to their survey and interview with practitioners \cite{humbatova2020taxonomy}.
\end{itemize}

\section{Variables}

In table \ref{table:var} we outline the variables we will measure for each issue. A detailed description of each variable is provided below.

\textit{Independent Variable:}
\begin{itemize}
%    \item \textbf{Initial components:} The software components mentioned in the initial issue report, this is the scripts and its related software modules that reporter include in the report. This contains information about the reporter's understanding of the root cause of the issue. In an ML projects, different components such as training pipeline, model training, monitoring, serving pipeline are interrelated, errors in one area can caused by defects in others components.
%    \item \textbf{Emerged components:} The newly mentioned scripts and its related software modules in the comments section of an issue report. This variable indicates the components that are not mentioned in the initial report that needs to be investigated. If the components which are already mentioned in the initial emerge in the comments, this indicate the critical components that require attention.
%    \item \textbf{Open-code topic:} The topics or categories of information of an issue report based on the issue title and the issue report description. This is generated from the open-coding process between 2 researchers who manually look at the issue. 
     \item \textbf{Category:} Issues will be categorised as either an ML issue (i.e., meets the filter criteria specified in \autoref{sec:collection}) or a non-ML (i.e., an issue that doesn't directly relate to ML, even though it occurs in the context of an ML project repository). ML issues will be further sub-categorised using the \textit{taxonomy of real faults in deep learning systems} \cite{humbatova2020taxonomy}.
%    \item \textbf{Description:} The text description of an issue, this is often used in previous literature to severity prediction and bug classification (CITE) and severity is correlated with resolution time (CITE).
%    \item \textbf{Initial topics:} The topics automatically generated by applying topic modelling technique on issue title and issue report description.
%    \item \textbf{Comment topics:} The topics generated by applying topic modelling on each sentence in the discussion comments of an issue report.
%    \item \textbf{Issue label:} Label of the issue tickets assigned manually by the projects owner or automatically assigned by bots, this is also known as the tag of an issue which is the first indication of the types of reported issue tickets. 
\end{itemize}

\textit{Dependent Variables:}
\begin{itemize}
    \item \textbf{Resolution time}: Measured by the number of days between opening the issue and marking the issue as closed (this operationalisation is chosen due to the simplictic nature of Github's issue tracking system, which does not have an explicit resolved state). Open issues, and issues closed without an associated PR (e.g., won't-fix) will be excluded from the study.
    %How much time does it estimated to take for an issue to be resolved based on the issue text and the comments for information exchange in the discussion comments.
    \item \textbf{Size of fix:} Measured by the number of lines of code changed in the PR(s) associated with the issue. Specifically, the number of inserted lines plus the number of removed lines. For consistency with the metrics reported by version control tools (git), we included comment lines as well as blank lines of code, and attempt to detect file renames to avoid overestimating the number of changed lines when a file is renamed.
    %In an ML project, there are more non-ML issues than ML issues being reported. We control the severity of issues to cut down the number of non-ML issues to create a balanced dataset.
\end{itemize}

\textit{Confounding Factor:}
\begin{itemize}
    \item \textbf{Project repository}: Github contains projects of different sizes and project maturity, and these project factors could affect the resolution time and size of fix. To prevent confounding, we sample an equal number of ML and non-ML issues from each project repository, thus ensuring that ML and non-ML issue sets are identically distributed in terms of project factors.
\end{itemize}

% \usepackage{booktabs}
% \usepackage{booktabs}
\begin{table*}[]
\begin{tabular}{@{}llllll@{}}
\toprule
\textbf{Class} & \textbf{Name}      & \textbf{Abb} & \textbf{Description}                                                                                                                            & \textbf{Scale} & \textbf{Source} \\ \midrule
Independent    & Category     & C           & Non-ML or ML issue (and subcategory if ML)                                                                             & Nominal        & Closed Coding                  \\
Dependent      & Resolution time    & T            & Number of days to resolve (close) the issue                                                                                   & Numeric        & Github API                  \\
               & Size of fix  & S           & Lines of code change required to fix the issue                                                                            & Numeric        & Github API                  \\
Confounding    & Project repository           & P            &  Repository the issue belongs to                                                                                                       & Nominal   &  Github API                           \\ \bottomrule
% Confounding    & Severity           & S            & Lines of code required to fix the defect reports                                                                                                        & Numeric   &                             \\ \bottomrule
%               & Maturity           & M            & The development state of the projects                                                                                                           & Likert Scale   &                             \\ \bottomrule
\end{tabular}
\caption{The variables to be measured for each issue }
\label{table:var}

\end{table*}

% Confounding Variable(s) and how their effect will be controlled (e.g., species type (Vulcan, Human, Tribble) might be a confounding factor; we control for it by separating our sample additionally into Human/Non-Human and using an ANOVA (normal distribution) or Friedman (non-normal distribution) to distill its effect).
\section{Dataset}

\subsection{Data Source}
In 2020, Microsoft released a paper \cite{gonzalez2020state} containing all AI and ML relevant projects from Github in the past 10 years. The dataset from this paper is used as a starting point, it contains 700 AI \& ML frameworks \& libraries, and 4,524 repositories using applied AI \& ML. 

\subsection{Collecting ML and Non-ML Issues From Open-source Projects}
\label{sec:collection}
\begin{figure}
  \includegraphics[width=\linewidth]{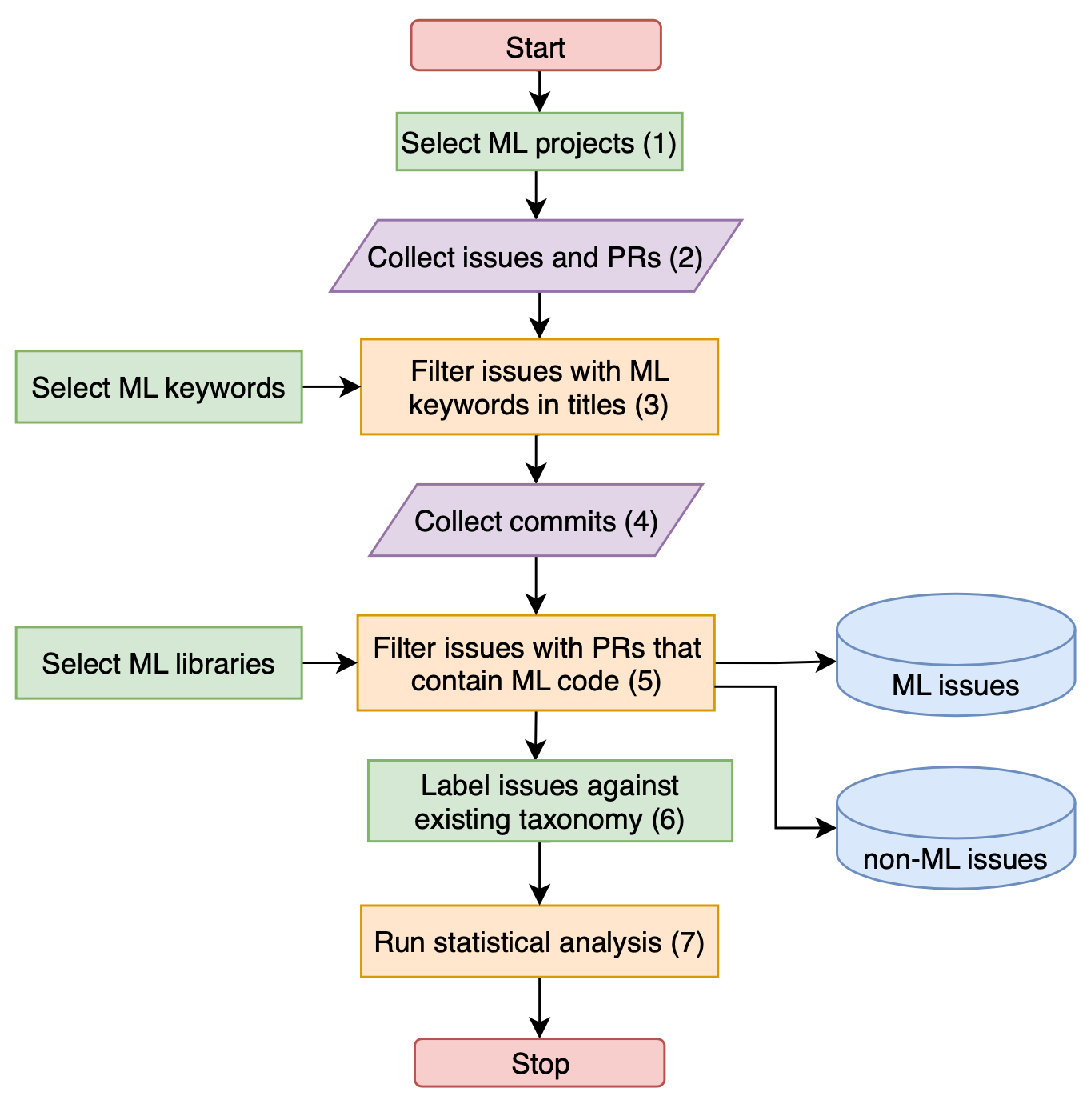}
  \caption{High level methodology for data collection and analysis (green: manual process, yellow: automatic approach, blue: dataset, purple: automatic approach using Github API)}
  \label{fig:methodology}
\end{figure}

To test all hypotheses the following method is proposed as shown in figure \ref{fig:methodology}:

\begin{enumerate}
    \item \textbf{Select ML projects}: We first randomly sample projects written in python from the list of ML projects to manually assess the data source. We chose to focus on python because it most popular languages for open-source ML projects on GitHub  \cite{Braiek2018}, ensures size of commits for our research questions are not confounded by different languages, and because the existing taxonomy \cite{humbatova2020taxonomy} that we will be comparing to was derived from artifacts and discussion related to deep learning frameworks for python. After that, we narrow down the number of projects for the manual categorisation of ML issues. We select ML repositories using the following criteria: newness (the projects must be created in the last 5 years), popularity (the projects must have at least 100 stars), activeness (the projects must have at least a commit in the last 2 years), and language (the projects must be written in python).
    
    \item \textbf{Collect issues and PRs:} We use Github API to get all closed issues of the selected projects from the previous step then map the PR that closed the issue.
    
    \item \textbf{Filter issues with ML keywords in titles:} We first start to \textbf{select ML keywords} from a subset of the ML Glossary from Google\footnote{https://developers.google.com/machine-learning/glossary}. The ML keywords glossary includes overloaded words such as "class", and "test" which could refer to either ML or non-ML concepts and thus lead to false positives. To mitigate this issue, we manually check that the filtered list of issues are using those words in an ML context. 
    
    \item \textbf{Collect commits:} For each PR, we use Github API to retrieve all commits associated with it and download the source code of each commit.
    
    \item \textbf{Filter issues with PRs that contain ML code:} We first \textbf{select ML libraries} currently used in popular python deep learning frameworks: tensorFlow, keras, pyTorch.
    After that, we filter issues with associated PRs in which the commits contain at least a python script with one of the selected ML libraries as an import statement. This will give us a \textbf{ML issues dataset}. This dataset satisfies two criteria: 1) issues contains ML keywords in the title, and 2) the fix for the issue is in the context of a script that uses ML libraries.
     
    We also sample a \textbf{non-ML issues dataset} from each selected project, in equal proportion to the number of ML issues detected in that project (this sampling strategy helps to prevent confounding effects arising from project factors). Non-ML issues are those that don't meet the ML criteria, but meet all other criteria (i.e., closed issue with an associated PR request).
    
    \item \textbf{Label issues against existing taxonomy:} We manually label ML issues in the ML issue dataset against an existing taxonomy \cite{humbatova2020taxonomy} using an iterative approach described in section \ref{sec:rq1execution}.
    
    \item \textbf{Run statistical analysis:} When there are more non-ML issues than ML issues or vice-versa, we will randomly sample the group with a greater number of issues such that the number of issues in each group is equal. After that, we compute the size of fix (variable S) and the resolution time (variable T) of each issue. After that, we examine how the category (variable C) of issue affects S and T using the visualisation and statistical tests described in section \ref{section:rq2execution} and \ref{section:rq3execution}.

\end{enumerate}

\section{Execution Plan}

\subsection{RQ1: What is the frequency of ML issue categories in open-source projects?}
\label{sec:rq1execution}

% The open-codes will be compared to information needed for traditional software defects specified in the literature to determine if the initial defect reports contain information unique to machine learning. 

We plan to manually label all the ML issues using an existing taxonomy of ML faults \cite{humbatova2020taxonomy}, and count the number of issues in each of the categories. We choose this taxonomy to follow because the research is published recently in 2020 in a well-established conference, is comprehensive using a mix method of survey, open-coding techniques, and it is empirically derived from real-world projects and discussion.
%Our effort will be the first to replicate and reevaluate their research, and will strengthen our analysis for the next research questions.

In our study, we test the reliability with which the taxonomy can be applied as an initial step. We follow an iterative process to validate the result of our classification against existing classes of bugs in the taxonomy. First, 3 raters study the taxonomy and the definitions of each category. After that, we randomly sample 30 issues and have 3 raters independently label them. The inter-rater agreement of the labelling process is measured by the Light’s Kappa metric \cite{light1971measures}, which equals the average of all possible combinations of bivariate Kappas between raters. This metric indicates an overall index of agreement. A low Light’s Kappa metric will be addressed by refining the taxonomy and repeating the process until the agreement is substantial (Kappa score above 0.61 \cite{landis1977measurement}). The first author will then independently label the rest of the issue dataset. For issues that do not fit in any category in the taxonomy, we put them in a separate list and open-code it later.

Finally, we compare the distribution of the issues  with the distribution from the existing taxonomy paper using descriptive statistics. 
Specifically, we will compare the percentage of issues in each category from labelling our dataset to the percentage of issues/posts/interviews that were used to construct each category in the existing taxonomy by \citet{humbatova2020taxonomy}.
 
%We will visualise the distribution via a grouped bar chart, each group in the bar chart represents the top level categories due to the small number of issues at the lowest level of categories in the taxonomy. Each group has 3 columns: Number of issues in each group from our manual process and the 2 numbers from the existing taxonomy (the number of posts assigned to such a category during manual labelling and the number of occurrences of such a tag in the interviews after their open-coding process).

\subsection{RQ2. How does the distribution of ML and non-ML issues compared in terms of resolution time and size of fix?}
\label{section:rq2execution}

To compare the resolution time distribution between ML and non-ML issues, we first calculate the resolution time for each issue as the difference in time between when the issue was first opened and when it was last closed. We only investigate issues that have an associated PR, which helps avoid inclusion of issues that were closed without being fixed. We compute distribution for ML and non-ML issues separately, then compare the two distributions. We plan to visualise the two distributions using a cumulative distribution plot, showing the percentage of resolved issues as a function resolution time. After that, we conduct non-parametric statistical tests to check whether and how the distributions differ; specifically a \mbox{K-S} test to determine if the shape of the distributions differ, and a Kruskal-Wallis test to determine if the medians differ.

We will repeat the above analysis procedure to study differences in the size of fix between ML and non-ML issues. Specifically, we measure the size of fix as the number of changed lines in the PR(s) associated with the issue. This analysis will add additional insight into any differences identified in resolution time. For example, if ML fixes are found to be of equal size to non-ML fixes, but still take longer to resolve, this would be suggestive that ML issues may take longer due to missing information in the defect report needed to efficiently locate and resolve the ML issue rather than being due to implementation complexity.

%Size of fix (put something in threat to validity): number of line being made to fix, remove whitespace line.

% Use survival plot, non-parametric test
% To answer this research questions, we take all the issues in the dataset D, measure the proportion of issues that still opened at time t (measure by days), and visualise using the survival plot.

%Do the test twice for size of fix and res time.
% For each issues in the dataset D, we use topic modeling on each comments in the discussion thread to get the categories of information being reported overtime.

% The time taken for an issue to be resolved is recorded and divided into 3 group using quartile interval: Micro time to fix, medium time to fix, and long time to fix. Each groups are equally one third of the total number of issues in the dataset D.

% From the list of categories of reported information in each issue and the 3 groups in terms of time to fix. We draw a table to compare, each cell is calculated by the following formula: Number of issues within the group in terms of time to fix with the information/ total number of issue in the group in terms of time to fix.

% We expect the group of defects required long time to fix have low percentage compared to the group with small time to fix.

\subsection{RQ3. How does the distribution of different ML issue categories compare in terms of resolution time and size of fix?}
\label{section:rq3execution}

To answer this research question, we compare the distribution of resolution time and size of fix for each of the issue categories labelled using the taxonomy of ML faults \cite{humbatova2020taxonomy}. We begin by comparing the distribution of resolution time and size of fix for each of the top level categories of issues in the taxonomy (model, tensors \& inputs, training, GPU usage, API). The Kruskal-Wallis test (a non-parametric version of ANOVA) will be used to determine if the medians of the groups differ. If there the test indicates significant differences at the top level categories, for each top level category, we will perform a test of whether there are differences between its subcategories. We will continue in this manner and stop descending branches of the taxonomy for which no significant differences are found between its subcategories, which could indicate either insufficient data at that level of granularity to explore further, or negligible differences between those categories in terms of resolution time and size of fix.

As this analysis involves a large number of statistical tests, Bonferroni correction will be applied to account for the number of tests performed at each depth level.

The outcome of this analysis will be an understanding of whether there are ML issue (sub)categories with larger resolution times and/or fix sizes than other ML issues, and are thus candidates for further investigation by future studies to determine potential interventions that could help resolve these (sub)categories of ML issues more efficiently (i.e., faster or using less lines of code).

\section{Threats to Validity}

% Group into Internal external construct validity

\subsection{Internal Validity}

The size of fix may be misleading in some cases, such as a project refactor. In particular, past research identified cases of repositories that encoded data as python files, leading to source files with over 50,000 lines of code \cite{Simmons2020}. To lessen the impact of this risk, our execution plan makes use of non-parametric tests (e.g., that test for a difference of the median rather than mean, and are thus less sensitive to outliers). Furthermore, our analysis involves visual inspection of the distributions which will allow identifying outliers and investigating possible causes.

\subsection{External Validity}

Our analysis assumes that ML issues are reported using the issue tracker, that the PR fixing the issue is linked to the relevant defect report, and that the issue is closed after resolution. However, ML development may follow a less formal process that traditional software, for example, may be conducted in notebooks using informal versioning practices \cite{Rule2018}, in which case ML issues with linked PRs containing fixes may not be present in the issue tracker. If our data collection results in few ML issue reports, future work will be to investigate whether ML issues are being reported and fixed through informal processes.

The scope of our study is limited to ML projects written in python. This was necessary so that we could identify ML code modules in a consistent manner (based on the libraries they import). Python is the most popular languages for open-source ML projects on Github \cite{Braiek2018}, but restricting the scope to a single language may still limit the generality of our findings.

Furthermore, we only analyse open-source projects, thus may not capture the issues faced by companies running proprietary ML pipelines in production. In particular, ML issues such as data shift are inevitable for companies running ML in production, but may not occur in open-source ML projects that serve as a library or are demonstrated on static datasets.

\subsection{Construct Validity}

Our study measures resolution time and fix size. However, the interpretation of these constructs in nuanced. A long resolution time could be because the issue was difficult to resolve, or it could simply be because the issue was considered low priority by developers. Similarly, an issue that is of a large size may indicate that it was more complex to resolve, or could simply point to poor coding practices.

%In the dataset we use, some ML issues might not be reported in the issue tracking system.

%Why we continue the study when the 1st RQ turns out there is no different between ML and non-ML?

%We only consider python script, bad for generalisation

%run time errors are hard to find in opensource

% \begin{figure}
%   \includegraphics[width=\linewidth]{figures/metamodel_process_updated.pdf}
%   \caption{High level methodology for selecting, labelling, and analysing projects}
%   \label{fig:metamodel-process}
% \end{figure}

\bibliographystyle{ACM-Reference-Format}
% \bibliography{references, specific-references}
%\bibliography{references}
\bibliography{specific-references}
% \appendix
% \input{appendix.tex}
\end{document}